# Secure Localization in Wireless Sensor Networks: A Survey

Waleed Ammar, Ahmed ElDawy, and Moustafa Youssef
{ammar.w, aseldawy, moustafa}@alex.edu.eg
Computer and Systems Engineering Department
Alexandria University
Egypt

July 7, 2009

## Abstract

Wireless sensor networks (WSNs) have gained researchers' attention in the last several years. Small sensors powered by miniaturized microprocessors are capable of supporting several applications for civil and military domains. Determining the location of sensors is a basic and essential knowledge for most WSN algorithms and protocols including data tagging, routing, node identification, among others. This paper surveys the different algorithms that have been proposed to securely determine the location of a sensor node. By "secure", we mean that adversaries cannot easily affect the accuracy of the localized sensors. In other words, the localization algorithm must be robust under several attacks.

We provide a taxonomy for classifying different secure localization schemes and describe possible attacks that can harm localization. In addition, we survey different secure localization schemes and show how they map to the proposed taxonomy. We also give a comparison between the different schemes, showing the attacks addressed by each.

## 1 Introduction

In a matter of just few years, WSNs have gained attention from researchers in different fields. A WSN consists of a large number of small sensors cooperating to achieve one goal. This helps leveraging several military and civil applications. Since these micro-sensors have limited power and computation, we need special algorithms with low power consumption. WSNs are usually deployed in harsh environments (e.g. battlefields) and are operating unattended. This makes them more vulnerable to adversary attacks.

Knowing a sensor's location is essential for many sensor network applications including surveillance networks and habitat monitoring. Equipping each sensor with a GPS device is too expensive because the number of sensors in a WSN is usually in the order of thousands.

There are several localization schemes [1] that allow sensors to determine their physical locations in absence of special hardware. Many localization schemes (anchor-based) assume that some special sensors (anchors) know their true physical locations. Other sensors determine an approximate relative location to anchors based on some measurements. However, there are some other schemes (anchor-free) in which there are no anchor nodes (e.g. [2]). In these schemes, sensors locations are calculated according to some virtual coordinates.

We expect that an adversary will try to prevent localization algorithms from working correctly. An adversary may inject malicious data into the network in order to displace sensor nodes. This means that when sensors estimate their locations, a displaced location is calculated. In this paper we survey different algorithms that can fall under such adversary attacks and still get a good estimate of sensors' locations.

In the balance of this section, we introduce some terminology that we will use in the rest of the paper and present the paper organization.

## 1.1 Terminology

*Anchor*: A node that knows its location without the aid of any other nodes. e.g. using GPS.

*Beacon*: A message sent by an anchor node that helps other nodes to know their locations.

*Prover (Claimant)*: A node that claims to be at some location. Other nodes in the network try to know whether this claimed location is correct.

$p$: A legitimate prover.

$p_m$: A malicious prover.

*Verifier*: A node among a group of nodes (verifiers) that tries to verify the location of another node (prover).

$v$: A legitimate verifier.

$v_m$: A malicious verifier.

$V$: A set of verifiers.

*Location verification*: The process in which a group of verifiers try to verify the location of a prover.

*Distance bounding*: A protocol run by a verifier to get an upper bound of the distance between it and a prover.

*Base station*: The gateway connecting the wireless sensor network to the operator.

## 1.2 Paper Organization

The paper is organized as follows. In Section 2, we provide a taxonomy for classifying different secure localization algorithms. In Section 3, we discuss attack models on localization schemes. In Section 4, different secure localization schemes are surveyed, then discussed in Section 5. In Section 6, we conclude the paper and discuss open issues.

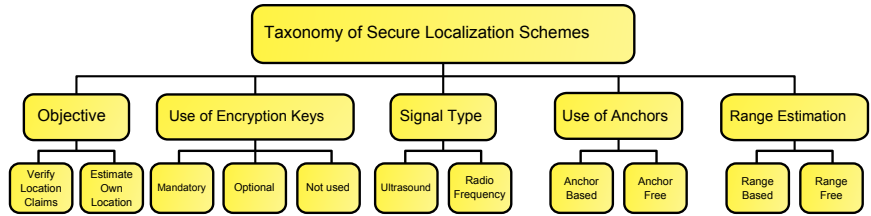

Figure 1: Classifications of secure localization schemes

## 2 Taxonomy

In this section, we present different criteria that can be used to classify secure localization algorithms in wireless sensor networks. This classification (Figure 1) reflects many design choices for such algorithms. This taxonomy is used later to classify all the algorithms we survey in this paper as shown in Table 1.

### 2.1 Objective

Secure localization algorithms can be classified into two main categories according to their objective. The first category aims to verify location claims of nodes. This is important because, usually, nodes are granted more services, privileges, trust or credibility when they are closer to other nodes (e.g. access the gateway's resources when within reach of it). So, an adversary node probably would claim to be in a closer location to an authority node than it actually is.

The second class of algorithms (location estimation) aims to assure that nodes can estimate their own true locations with the help of other nodes, even when some of the nodes are malicious. For example, if adversary nodes could deceive legitimate nodes, making them estimate their locations inaccurately, adversary nodes can exploit location-based routing algorithms to modify, repeat, or drop all messages passing through them. This is one reason why robust own location estimation is an important problem to address.

### 2.2 Use of Encryption Keys

There are four types of encryption keys that are likely to co-exist in a wireless sensor network [3]:

- An individual key shared with the base station

- A pairwise key shared with another node

- A cluster key shared with multiple neighboring nodes

- A group key shared by all the nodes in the network

Establishment of those keys in a network, and the usage of each are beyond the scope of this survey, interested readers are referred to [3].

There are three classes of secure localization algorithms in regard to the use of encryption keys: First, algorithms that are based on encryption keys. Such algorithms cannot be implemented unless encryption keys are set up in the network. Second, algorithms that can make use of keys if they exist, but still can operate in their absence. Third, algorithms that do not make use of keys, even if they happen to exist in the network.

The use of encryption keys makes an algorithm more resilient to attacks. But this comes with a price, that such algorithms are less feasible due to the computation overhead of establishing encryption keys.

## 2.3 Signal Type

Two types of frequencies have traditionally been used in secure localization: ultrasound and radio-frequency (RF). Ultrasound is considerably slower, which is good in the sense that location estimation is not sensitive to errors in time estimation. But, being too slow (with respect to RF) means that two (or more) adversaries may communicate using RF to shorten the estimated distance. Ultrasound propagates in metals faster than in air, allowing for some physically-present attacks [4]. RF signal is preferred because it moves through objects. This is a desirable property, for larger coverage, especially when the network is deployed randomly.

## 2.4 Use of Anchors

In anchor-based algorithms, a small percentage of nodes, namely "anchors", know their absolute locations, usually using a GPS receiver. The algorithm then tries to estimate the absolute location of other nodes in the network, using the location information provided by the anchors in beacon frames. In contrast, anchor-free algorithms try to estimate relative locations of nodes in the network without making any assumptions about nodes' absolute locations. Anchors make calculations more accurate. On the other hand, anchor-free algorithms are more scalable because they do not require special types of nodes. Note that a virtual coordinate system constructed using an anchor-free algorithm can be mapped to an absolute coordinate system by getting the locations of three non-colinear nodes in case of 2D (four in case of 3D).

## 2.5 Range estimation

Localization algorithms can be generally classified into range-based and range-free algorithms [5]. Range-based algorithms use absolute point-to-point range estimates (e.g. distance or angle) for estimating location [1]. Range-free algorithms employ other methods to approach the localization problems (e.g. using the hop-count between two nodes as an estimate of the distance between them). Range-based algorithms are generally more accurate because they use physical quantities in measuring distances more accurately.

# 3 Attacks

Before getting into the details of each algorithm, it is useful to know what attacks could harm the localization schemes in wireless sensor networks.

## 3.1 Distance fraud attack

A distance fraud attack is based on hacking the method of measuring distance between two nodes. It consists of one malicious node trying to appear closer or further to another legitimate node. This attack could take different forms:

- When a verifier node sends a signal to a prover node, the prover node is supposed to send the response signal immediately after receiving the verifier's signal. A malicious prover could send the response even before receiving the verifier's signal to appear closer. A malicious prover could also introduce delay before sending the response to appear further away from its actual location [6].

- Measuring signal strength is one of the methods of estimating distance between nodes. All nodes are supposed to use the same power level for transmission. A malicious node, however, could increase the power level for transmission so that a receiver measures a stronger signal, indicating the malicious node is closer than it really is.

- Measuring signal strength can be used in another way: a verifier node sends out a message with a particular signal strength, making the transmission range equal to the desired distance. Only nodes within this distance can hear the message sent by the verifier. A malicious prover still can increase its receiving range using a high gain or directional antenna [7].

- When ultra-sound signals are being used to measure distance between nodes, physically present attackers could place metallic objects between the two nodes (since the sound travels faster in metals than in air). Hence, the two nodes would appear closer to each other than they really are [7].

## 3.2 Mafia fraud attack

A Mafia fraud attack [8] is a man-in-the-middle attack [9]. It consists of a legitimate prover $p$, a legitimate verifier $v$, a malicious prover $p_m$, and a malicious verifier $v_m$ (see Figure 2, based on [7]). When $p$ is about to perform the verification protocol with $v_m$, the latter establishes a radio link with $p_m$ and sends any information transmitted by $p$ straight to $p_m$, which in turn sends it to $v$. When $v$ tries to verify the identity of $p_m$, same thing happens in reverse order [6]. This causes the distance between $p$ and $v$ to be calculated incorrectly.

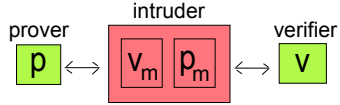

Figure 2: Mafia fraud attack

## 3.3 Terrorist fraud attack

A terrorist fraud attack is similar to a mafia fraud attack, except that the prover is maliciously collaborating with the intruder to fool the verifier. The dishonest prover uses the intruder to convince the verifier that it is close [7].

## 3.4 Wormhole attack

In the wormhole attack, a malicious node records packets at one location in the wireless sensor network, tunnels them to another malicious node at another location, and retransmits them there to distant sensor nodes. This would make these sensor nodes estimate their locations incorrectly [10]. Even when public-key encryption is used to authenticate nodes, two malicious nodes could disclose their private keys to each other [7].

## 3.5 Sybil attack

In anchor-based localization schemes, if an adversary compromises an anchor node (which is supposed to help other sensor nodes determine their locations) then the adversary can pretend to be in two (or even more) different locations, confusing nearby sensor nodes [11].

## 3.6 Spoofing attack

A spoofing attack consists of an adversary setting up a malicious node to provide false localization information by pretending to be a legitimate node in the network.

## 3.7 Jamming

An adversary broadcasts a high-energy signal to disrupt network operation [12].

## 3.8 Overshadowing

The original message appears as noise in the adversary's much stronger signal [13].

## 3.9 Manipulation

An adversary modifies the content of messages that contain localization information.

## 3.10 Replay

An adversary plays back a past legitimate localization message to same/other recipients.

# 4 A survey of secure localization algorithms

The algorithms covered in our survey are grouped according to their objective, to make it easier for the reader to understand the point of each algorithm. In Section 4.1 we survey algorithms used by a group of sensor nodes to verify the location of another node. In Section 4.2 we survey algorithms used by a sensor node to estimate its location correctly.

## 4.1 Verifying location claims

The protocols described under this category aim to verify location claims of nodes. This is particularly important because, usually, nodes are granted more services, privileges, trust or credibility when they are closer to other nodes. So, an adversary node would probably try to claim a closer location to an authority node.

### 4.1.1 The distance-bounding protocol

By introducing this protocol, Brands and Chaum [6] do not solve the problem of verifying location claims. Rather, they solve the closely related problem of a verifier that wants to assure that a prover is within a certain distance. The distance-bounding protocol has been extended in so many ways in wireless localization schemes [7, 10, 14, 15]. A verifier sends out a single-bit challenge. A prover, immediately after receiving the challenge, sends out a single-bit response. The verifier logs the round-trip time (the time it took the response to reach the verifier after sending out the challenge). This process repeats K times (K being a security parameter). After that, the verifier determines the upper-bound on the distance by multiplying the maximum delay (i.e. the maximum round-trip time) by the known signal propagation speed.

Brands and Chaum showed how this can prevent a *mafia fraud attack* described in Section 3.2. Recall that the main players in a mafia fraud attack are: a legitimate prover $p$, a legitimate verifier $v$, a malicious prover $p_m$, and a malicious verifier $v_m$ [7]. The $K$ challenges and the $K$ responses are chosen at random, and the rapid bit transfer is performed (i.e. sending the $K$ challenges and receiving the $K$ responses). $p$ signs, using its secret key, the concatenation of all the $2k$ bits and sends the signature to $v$, which accepts it if and only if the received signature is correct, and finds the upper bound on the distance using the rapid bit transfer. There is a possibility that $p_m$ guesses the random responses of $p$. However, the possibility decreases exponentially with increasing the value of $K$.

This leads to another question, away from the mafia fraud attack: what if the prover $p$ itself tricks a verifier $v$ by sending its responses before receiving $v$'s challenges? The first solution is for $v$ to randomly select the time when to send its challenge bits so that $p$ does not know when exactly to send the corresponding response bits. When the verifier uses only two different random times, the probability of a successful attack becomes $2^{-K}$. The second solution suggests that response bits should be dependent on their respective challenge bits in one way or another. For example, the verifier may send a nonce with each challenge and requires the prover to send it back with its response.

### 4.1.2 SPINE

In [16] Čapkun and Hubaux developed a system for verifying the location of a wireless device in a network: Secure Positioning In sensor NEtworks (SPINE). Multiple nodes in the network, *verifiers*, run the distance-bounding algorithm to get a verified distance to the required node, *claimant*. These verified distances are collected at some node, *authority*. The location of the claimant is estimated using Minimum Mean Square Error (MMSE) based on bounded distances.

Since RF-based distance measuring techniques are based on EM (Electro Magnetic) waves, which travels at the speed of light, an attacker cannot reduce a measured distance. However, an attacker can still enlarge the distance. In order to detect distance enlargements, the claimant has to be located within the triangle formed by three or more verifiers. To verify the claimant's location, the authority node compares each bounded distance to the corresponding estimated distance (i.e. distance between the MMSE estimated location of the claimant and a verifier). If the difference between bounded and estimated distance is more than a predefined threshold, the estimated location is rejected.

To find the correct location, when the estimated location is rejected, the authority node performs the following computations. Verifiers are split into two sets: $\mathcal{C}$: verifiers with correctly measured bounds, and $\mathcal{NC}$: verifiers whose bounds are suspicious. $\mathcal{C}$ is filled with verifiers that form at least one verification triangle which contains $u$ (the claimant) and the estimated position of $u$ is correct in this triangle. The correct position in a triangle is calculated by trilateration using distance bounds of the head nodes. After that, new position estimation is calculated using verifiers in $\mathcal{C}$ with MMSE.

### 4.1.3 Location verification using secure distance-bounding protocols

In this protocol, Singelee and Preneel [7] modify the distance-bounding protocol [6] to be useful for authenticating users given their locations (rather than using passwords or smartcards). Unlike SPINE, this protocol uses encryption and introduces the "broadcast mode" of distance bounding algorithms (dicussed later in this section). To prevent the distance fraud attack, many solutions are available. First, a verifier can send out a nonce controlling its signal strength to achieve the desired transmission range. If the prover is within this range, it can prove it by sending back the nonce. Note that a prover may use a directional antenna to increase its sending or receiving ranges. So, this method is useful only when provers are guaranteed to have a standard unmodified device.

A second solution is to use a challenge-response approach, as in the distance-bounding protocol [6]. The response should be designed such that a prover cannot send it before receiving the challenge, and also does not involve heavy computations. The prover should use its private key somewhere in the scheme so that the verifier knows that the prover is the one replying (yet, this does not prevent wormhole attacks (Section 3.4) in which two attackers disclose their private keys to each other).

The distance-bounding protocol can solve the problem of mafia fraud attack, but not

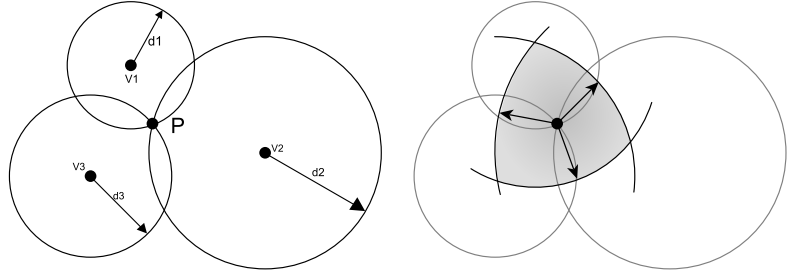

(a) Secure bounding protocol using 3 verifiers. Here we assume that the three distances are measured exactly.

(b) Secure bounding protocol when each distance is prolonged with the same value $d$. The lengths of the three arrows are the same value ($d$). The dark area is the expected region of the prover.

Figure 3: Geometric properties of broadcast mode [7]

terrorist fraud attacks, (Section 3) because in the latter, the in-the-middle node can perform the rapid bit-transfer and then send the response bits used to the prover so as to do the signing (which requires its private key) without sending the private key itself to the intruder node. The authors modified the distance-bounding protocol by engaging the private key in the rapid bit-transfer.

To know the exact location of the prover in a plane, we need three collaborating verifiers. The prover performs a distance bounding protocol with each of the three verifiers. If the distances found are exact, we can easily calculate the exact location of the prover (Figure 3(a)). If there is a delay, which is typical, the intersection of the three circles around the three verifiers (each having a radius equal to the verified upper bounded distance) will form an area where the prover is guaranteed to exist (Figure 3(b)). The verifiers may not accept the location claim if the area is too large because a malicious prover who wants to appear in a different location than its original location may delay the responses.

An attacker can pretend to be at a different location that is further away from all three verifiers. This can be achieved by carefully selecting the delay when performing the secure distance bounding protocol with each of the three verifiers. This is possible because the three instances of the secure distance bounding protocol are executed independently. In the "broadcast mode", only one instance of the secure distance bounding algorithm is used with the three verifiers simultaneously making sure that the intended (carefully calculated) delay is equal for all the three verifiers. The resulting area can be reduced to a point because the distance between the prover and the three segments of the area is the same (since the area was originally a point and kept expanding due to the delay, the expansion is equal at the three directions opposite to the three verifiers). In Figure 3(b), note that lengths of the three arrows are equal. This geometric property can be expressed analytically to solve for the exact location of the prover. Note that broadcast mode requires that all three verifiers are synchronized.

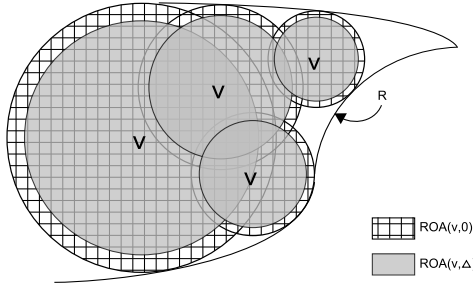

Figure 4: Combining the ROA for individual verifiers, the total coverage of the region of interest ($R$) increases [4].

### 4.1.4 Secure verification of location claims

In this protocol, Sastry *et al* [4] provide a solution for a set of verifiers ($V$) that want to make sure a prover ($p$) exists somewhere in a region $R$. It assumes that many trusted verifier nodes have secure communication between them. Note that any node within the region $R$ is considered legitimate (regarding what we need to verify). More than one malicious prover may coexist with a secure RF communication between them, as well as sound generation and detection capabilities.

With a single verifier ($v$) and a prover ($p$), $p$ sends its location to $v$. $v$ responds with a nonce using an RF signal, and registers the time of transmission. $p$, after processing time ($\leq \Delta_t$), responds with the same nonce using ultra sound. $v$ registers the time of receiving this signal. $v$ accepts the location claim $l$ iff $l \in R$ and elapsed time $\leq (d(v,l) - \Delta)/(\bar{c} + \bar{s})$[1]. Assuming the area $R$ is a circle centered at $v$ with radius $r$, the region of acceptance (ROA) (i.e. the area in which the verifier node $v$ is sure that it can correctly verify claims for a prover) is a circle centered at $v$ with a radius equal to $(d(v,l) - \Delta)$. Transmission time should also be taken into consideration as an overhead (i.e. added to the value of $\Delta_t$).

Figure 4 shows that using multiple verifiers, the ROA becomes more flexible and more probable to cover the region of interest ($R$). Moreover, if there exists pre-shared keys between the verifier and the prover, a challenge response mechanism replaces the nonce sending and receiving.

Note that a malicious prover would not add intended delay, since it will make the verifier think the prover is further than it is, which results in less services, privileges, trust or credibility. Yet, a malicious prover can exploit the difference in propagation speed of sound in different media. This allows an adversary to spread some metals in the field which causes sound to propagate faster, and hence, distances to be measured shorter. That is why it is advisable to measure the on-site propagation speed when the verifier is surrounded with metals. More than one verifier will do a better job verifying this location, but the attack is still possible though.

---

[1]$\Delta = \Delta_t.(\bar{c} + \bar{s})$, $\bar{c}$ = speed of RF signal, $\bar{s}$ = speed of sound.

### 4.1.5 SLA

Anjum *et al* proposed the SLA algorithm in [17] which is used to determine the location of a node. This algorithm is based on the ability of anchor nodes to vary their transmission power. This results in different transmission ranges for each anchor node. The intersections of these different transmission ranges divide the Area Of Interest (AOI) into different subregions (Figure 5).

First, a key $k$ is generated for each node. All anchor nodes are able to receive the keys securely, yet generating and distributing keys is not discussed in the paper. Each anchor node sends a set of nonces at different power levels using this session key. A sensor node to be localized should reply back the set of heard nonces for each anchor. Collecting this data at a sink node allows for determining the subregion at which the node under test lies. The sink node determines the location of nodes either using geometric properties (for the idealized case) or using a "message map" created beforehand which contains information about the set of messages that can be received at each location in the AOI (see Figure 5).

The accuracy of this algorithm is analyzed by calculating the average area of a subregion. It is calculated by dividing the AOI over the number of subregions. The average area of a subregion is found to decrease as the number of different power levels increases.

After that, the robustness of this algorithm against attacks is discussed. The threat model assumes that the sink node as well as anchor nodes are trusted and cannot be overtaken by an adversary. Only sensor nodes to be localized could be malicious. Malicious nodes are assumed to be normal nodes. This means they cannot move, cooperate with each other or have any specialized hardware like directional antennas. Under these constraints, a malicious node may only drop a set of nonces received, instead of replying back all received nonces. This allows a malicious node to claim that it is in another nearby subregion. For example, in Figure 5, a node in subregion 1 may claim it is in subregion 2 by dropping the first nonce received from $AP_2$. The authors proposed three techniques to limit such an attack. First, anchor nodes could use more power levels which will result in an increased number of subregions. Second, anchor nodes may send nonces at random. This makes it harder for the malicious node to know the power level of each nonce. Third, more anchor nodes may be deployed such that each point is covered by more that three anchor nodes.

Simulations showed that using a large number of power levels result in higher accuracy and robustness. The authors also compared using uniform power levels to non-uniform power levels. Uniform power levels mean that transmission ranges are $r, 2r, ..., kr$; $kr = R$, where $k$ is the number of different power levels and $R$ is the maximum transmission range. Results of simulations show that using non-uniform power levels is better in terms of robustness and accuracy.

Authors of the SLA algorithm showed, in a later publication, how to apply the same idea for verifying locations of static users in a WLAN [18].

### 4.1.6 Summary

We talked about verifying the location of a claimant. At its simplest form, the distance-bounding protocol [6] manages to put an upper bound on the distance between the verifier and the claimant. This allows to detect mafia fraud attacks.

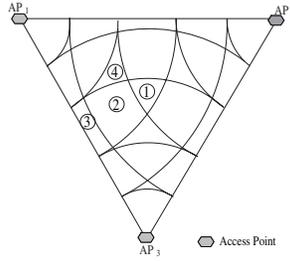

Figure 5: Message map when 5 transmission power levels are used by each AP. The regions marked 1, 2, 3 and 4 are different locations in the map [18].

As an extension to it, SPINE [16] uses the verifiable multilateration algorithm to verify location claims. In this algorithm different verifiers independently perform distance-bounding with the claimant. The bounded distances are then combined using MMSE and false bounds are detected and filtered out.

In [7] a more secure version of distance-bounding protocol is used by involving the private key in the rapid-bit-transfer. This allows to detect the terrorist fraud attack in which the claimant is maliciously collaborating with the intruder to fool the verifier. However, it is assumed in the attack model that the claimant is keeping its private key from the intruder; actually, this is not realistic. A broadcast mode of the distance-bounding protocol was used in which the algorithm is executed once and all the verifiers get their upper bounds. In this case, the error is the same in all bounded distances hence the region is reduced to a point.

In [4] RF is combined with ultra sound to get the distance between a claimant and the verifiers. This allows the verifiers to make sure that the claimant is within some specified region R.

In SLA [17] a node is localized by a set of anchor nodes. Anchor nodes send different messages with different power levels. The node replies back the set of nonces received and the sink can determine its region given the set of messages heared.

Table 1 gives a classification for the discussed algorithms.

## 4.2 Robust location estimation

The protocols described under this category aim to assure that nodes can estimate their own true locations accurately, with the help of other nodes. This is important to secure other algorithms such as routing. For example, if adversary nodes could deceive legitimate nodes, making them declare false locations for themselves, adversary nodes can exploit location-based routing algorithms to modify, repeat, or drop all messages passing through them.

### 4.2.1 SeRLoc

In [11], Lazos and Poovendran discuss Secure Range-Independent Localization Scheme (SeRLoc). SeRLoc is a two-tier localization scheme for WSNs. The network consists of sensors of unknown positions and locators aware of their positions (e.g. using GPS receivers). Sensors antennas are omni-directional while locators antennas are directed. Communication ranges of sensors and locators are well known before network deployment.

The basic idea of localization in SeRLoc is simple. A sensor receives location information from near locators. Based on this information, a sensor can define an intersection region in which it resides. Its location is estimated to be the centroid of the intersection region. To secure this simple localization method, authors identified three attacks (wormhole attack, sybil attack, and compromised entities) and showed how to address each of them.

**Wormhole attack**: An adversary can use the wormhole attack to tunnel beacons sent from one locator to a far sensor. This would cause this sensor to estimate its location incorrectly. To detect a wormhole there are two methods.

1. Single message per locator/sector property: If the sensor lies within the communication range of the tunneled locator, it will receive multiple messages from the same locator. Hence, a wormhole attack can be detected.

2. Communication range constraint property: Since a sensor may receive a beacon message from locators within a specified range, messages tunneled from a far locator can be detected. Assuming locator-to-sensor communication range $R$, the maximum distance between two locators sending to the same sensor is $2R$. When a sensor receives two messages from locators more than $2R$ apart, it detects a wormhole attack.

To recover from a wormhole attack, an algorithm called ACLA (*Attach to Closer Locator Algorithm*) is used. Simply, the sensor assumes that the first locator to reply is the closest and hence not tunneled. Any locator with a sector not intersecting with the first one is discarded.

**Sybil attack**: If an adversary managed to reveal locator IDs and keys, it can duplicate existing locators. However, it cannot add new locators since the sensors are preloaded with valid locator IDs. For the sybil attack to displace a sensor, it must clone a number of locators more than the valid ones in the communication range of this sensor. In this case, the sensor detects that the number of locators is too large compared to locator density ($\rho_L$) previously loaded to all sensors. When a sybil attack is detected, ACLA is used to resolve it.

**Compromised network entities**: When a locator is compromised it can displace nearby nodes. ACLA cannot solve this because, when a sensor node finds itself under attack and executes ACLA, the compromised locator might be the closest and the location may be displaced. In this case ELRA (Enhanced Location Resolution Algorithm) is used instead of ACLA.

For this scheme to be robust against such attacks, first, an authentication mechanism is used for sensor-locator communication. There is a pairwise key $K_s^{L_i}$ between a locator $L_i$ and a sensor $s$. These keys are preloaded at sensors and can be generated at locators.

The idea is that locators authorize themselves before sending beacon information to sensor. Since locators know their true position, it is easy for them to check when a compromised locator tries to cheat its location.

### 4.2.2 HiRLoc

As an improvement to the SeRLoc, Lazos and Poovendran introduced High-resolution Robust Localization for WSNs (HiRLoc) [19]. This algorithm is more accurate as each locator sends more information over time. Like SeRLoc, the locators antennas are directional. Unlike SeRLoc, communication range of locators is variable. Each locator sends beacon information multiple times (AKA rounds). In each round the locator may change either its antenna direction, its communication range or both. This allows the location determination process to be more accurate and robust.

The algorithm works as follows. First, sensor $s$ determines the locators of interest $LH_s$ as the locators heard by $s$ and it determines an initial estimate for its location. Then, it collects beacon information from locators in rounds. In these rounds, a locator may vary its direction, communication range or both. After that, a region of intersection (ROI) is calculated from the range estimates sent by all locators.

Defense against attacks is very similar to SeRLoc. However, range variation in HiRLoc is a challenge. When there is a *worm hole* attack and the range variation technique is applied, the detection mechanism used in SeRLoc cannot be used. To overcome this problem, ROI is calculated at the initial round. Then, any ROI of a later round that does not intersect with the initial one is assumed to be under attack.

**Sybil attack**: In antenna orientation variation, the same technique used in SeRLoc manages to detect the Sybil attack. In communication range variation, the same technique used with wormhole attack is used.

**Compromised network entity**: The same detection mechanism used in SeRLoc is used in HiRLoc.

As a conclusion HiRLoc is more accurate than SeRLoc but it incurs more communication and computation overhead.

### 4.2.3 Attack-resistant location estimation in sensor networks

In [20] Donggang *et al* present two algorithms to calculate a node's location using distances to other nodes. It is assumed that there is some method to calculate a distance between two nodes. Each node collects distances to other nodes. Some of these distances may be invalid due to an attack. Any of the two algorithms can be used to filter out invalid distances and calculate the location using benign ones. The algorithms used to detect invalid distances are *Attack-Resistant minimum mean square estimation* and *Voting-based location estimation*.

**Attack-Resistant minimum mean square estimation**

The idea of this algorithm is that invalid distances introduced among benign distance will introduce some inconsistency. Authors suggest that inconsistent distances are the malicious ones. To detect inconsistent distances, first, MMSE is used to calculate location estimation. Then, *mean square error* is calculated. If the error is above some threshold, an inconsistency is detected. Inconsistent distances are removed and the process is repeated.

Each anchor node located at $(x_i, y_i)$ sends a location reference $\langle x_i, y_i, \delta_i \rangle$ where $\delta_i$ is the distance measured from its beacon signal. When a set of location references are found to be inconsistent the algorithm tries to remove each single location from this set until a consistent subset is found. If all subsets are found to be inconsistent, the algorithm chooses the one with the minimum mean square error. After removing a suspected distance, the whole process is repeated.

**Voting-based location estimation**

Imagine that a node is "lost" and other nodes are trying to guess where it is. The search space is divided into a grid of small cells. A voting is made on each cell to determine its goodness to be the location of the lost node. The centroid of the cells with the highest votes is used as an estimate location. After calculating this estimate, location references that are too far from this estimate are detected as malicious and removed.

The search space is divided into square cells with a fixed side length $L$. $L$ could be a parameter to the system that determines the accuracy of the process. Each cell holds a vote counter initially set to 0. For each location reference $\langle x_i, y_i, \delta_i \rangle$ (which means that the node at $(x_i, y_i)$ thinks that the lost node is $\delta_i$ away), a *candidate ring* is defined as the ring centered at $(x_i, y_i)$ with inner radius of $\max\{\delta_i - \epsilon, 0\}$ and outer radius of $\delta_i + \epsilon$, where $\epsilon$ is the distance calculation error. Each cell that intersects with a candidate ring has its vote counter incremented. At the end, the centroid of the cells with the maximum vote counter is used as a location estimate.

Since this algorithm runs on sensor nodes with limited computation and power, authors developed a fast method to determine whether a cell intersects with a ring or not [21].

### 4.2.4 SecNav

SecNav [13] consists of a set of stations forming a navigation infrastructure which provides radio signals that enable devices to determine their location and to obtain an accurate time reference. Infrastructure stations are synchronized and carefully placed to cover a certain region (each point is within the range of at least four infrastructure stations). It is a broadcasting protocol that does not require the navigation devices (which want to locate themselves) to transmit any messages.

An attacker can relay or delay transmitted messages. But she cannot disable the channel (e.g. using a Faraday Cage [13]). She can eavesdrop, insert messages, or send jamming signals.

SecNav (Secure Broadcast Localization and Time Synchronization in Wireless Networks) assumes that infrastructure stations are secure against adversary compromises. The main idea of SecNav, as illustrated by Rasmussen *et al*, is to encode navigation signals using integrity-codes. To ensure the message will contain an equal number of ones and zeros, Manchester encoding is applied. Then, on-off keying is applied such that 1 corresponds to a random waveform, while 0 corresponds to silence. Note that an attacker cannot turn a 1 into 0 (since this requires removing the random signal from the channel). Figure 6 illustrates the integrity codes.

Jam, overshadow, forgery, manipulation, and replay attacks can be detected because an attacker may not alter a one into a zero. If she tries to alter a zero into a one, the ratio between ones and zeros shows that there is an attack. Overshadowing is detected because: for the receivers to detect symbols 0 and 1 on the channel, receivers measure strengths of the received signals (as opposed to their signal-to-noise ratio). An adversary may also set up a malicious infrastructure station that provides false location and time information. To avoid this, legitimate infrastructure stations should keep the channel busy by either sending valid navigation messages in an uninterrupted sequence or by transmitting I-coded (integrity codes) sequences.

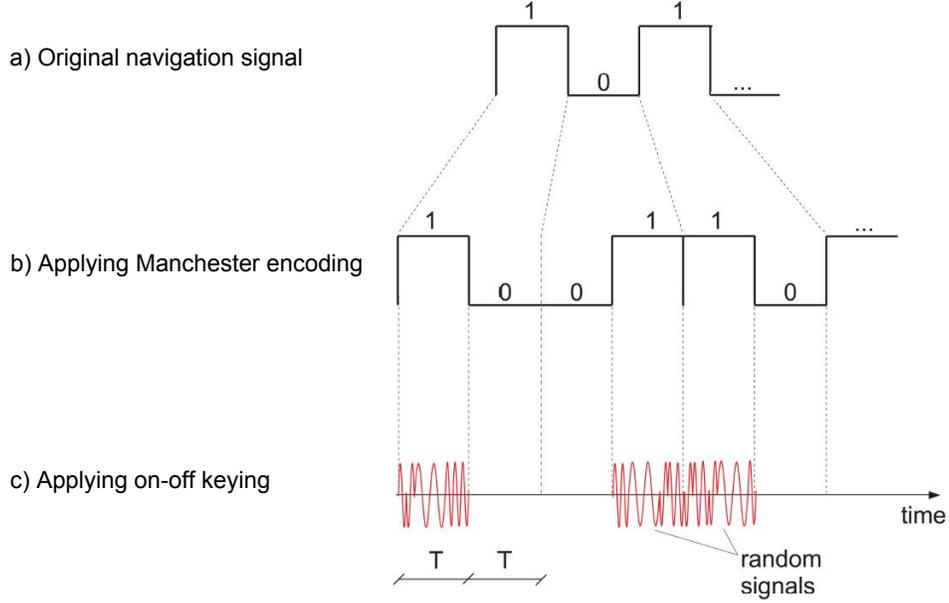

Figure 6: SecNav navigation message encoding [13].

### 4.2.5 ROPE

Lazos *et al* proposed ROPE (Robust Position Estimation in Wireless Sensor Networks) [14], which assumes a two-tier network topology consisting of:

- Sensors: randomly deployed with density $\sigma_s$ each with an omnidirectional antenna for sensor-sensor communication with range equal to $r$.

- Locators: randomly deployed with density $\sigma_L << \sigma_s$. Each is equipped with M directional antennas of beamwidth $2\Pi/M$ for locator-to-sensor communication with range $R > r$. Sensor-to-locator communication range $r_{sL}$ is larger than r due to the antenna directivity gain of the locators antennas. Each locator knows its location and orientation (e.g. using GPS). Each sensor shares a pairwise key with each of the locators $(K_{Li}^s)$

A sensor broadcasts its id $(Id_s)$ and a random nonce $(N_s)$. Any locator that can communicate bi-directionally with $s$ performs distance bounding with $s$. This assures that:

- The sensor is within the vicinity of the locator

- The sensor can define the $LDB_s$ set as

$$LDB_s = \{L_i : distance(L_i, s) < r_{sL}\}$$

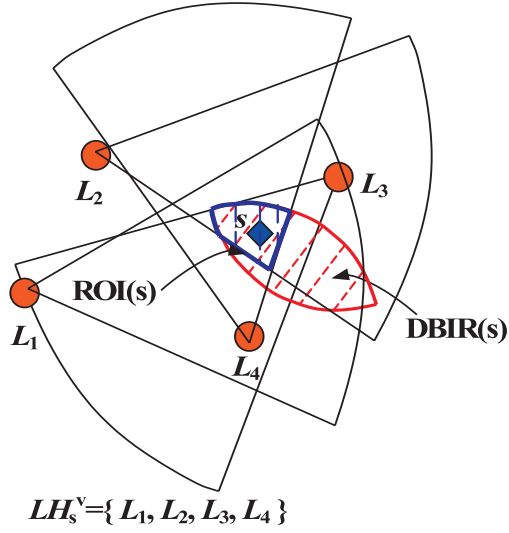

Figure 7: DBIR(s) is the intersection of all $D_i s$.

If $|LDBs| > 3$, and $s$ lies inside a triangle formed by three locators $L_i \in LDB_s$, $s$ computes its location using verifiable multilateration. $s$ sends a termination notification to $L_i$'s that can hear it. Else, for each $L_i$ in $LDB_s$, $s$ computes the disc $D_i$ according to the distance bound and given the orientation and antenna direction of each $L_i$. $s$ then computes the Distance Bounding Intersection Region DBIR(s) as the intersection of all $D_i s$ (see Figure 7). If locator $L_i$ in $LDB_s$ has not received a termination notification, it re-broadcasts the initial sensor message and its own $Id_{Li}$. This broadcast covers all locators at distance $\leq R$ from $s$. Each locator $L_j$ that hears this re-broadcast will transmit at each sector (i.e. on each directional antennas) a message and its MAC (using its pairwise key with $s$). The message contains the locator Id, its location and orientation, and the original nonce sent by $s$ ($N_s$) to guarantee freshness. $s$ receives the latter transmission from a set of locators $LH_s$. $s$ defines ROI (region of interest) as the region in DBIR(s) where most sectors of $LH_s$ intersect.

Location Verification is simply done between $s$ and any (one or more) locators close enough (i.e. $\|L_i s\| < r_{sLi}$), using distance bounding.

### 4.2.6 DRBTS

In this work, Srinivasan *et al* tries to answer the question "Given a network of anchors and sensors, how to exclude malicious anchors that provide sensors with incorrect location information?" Distributed Reputation-based Beacon Trust System (DRBTS) uses the concept of reputation for excluding anchor nodes [22]. Every anchor node maintains a Neighbor-Reputation-Table (NRT). An anchor node monitors its 1-hop neighborhood for misbehaving anchor nodes, and update its NRT accordingly. Then, it publishes the NRT enabling other anchor nodes to update their own NRTs, and enabling nearby sensor nodes to determine whether or not to use a given anchor's location information.

The authors discussed a number of assumptions, most importantly:

- Location information is broadcast to the requesting sensor node by the anchors.

- Location information is encrypted using a networkwide group key.

- The majority of anchors are honest.

To initialize an anchor's reputation, when an anchor hears another anchor responding for the first time, it sets the new anchor's reputation to 0 before evaluating the transmission. The drawback of this sort of initialization is that no one trusts anyone at the beginning; disabling the whole localization capability of anchors. To bootstrap the system, each anchor is given a small number of fake IDs to distinguish itself as a sensor node and request location information, triggering responses from nearby anchors; allowing reputation to build up. To speed up building of anchor reputations, neighboring anchors are allowed to share their experiences. Publishing reputation information is associated with dissemination of location information, mapping the publish-rate to the rate at which reputation changes.

Each anchor node waits for location inqueries from sensor nodes in its range, and responds with its location, as do all other anchor nodes within the range of the requesting node. In addition to the location, an anchor node also responds with its reputation values for each of its neighboring anchors. Other anchors within the 1-hop neighborhood will evaluate this response in light of their own locations and NRTs using a deviation test. Meanwhile, the sensors will also receive the anchors' responses, and use them to decide whether or not to use a given anchor's location information, based on a simple majority voting scheme. If one anchor responds showing it trusts another anchor, the sensor counts that as a positive vote from the first anchor to the second. For an anchor to be trusted by a sensor, it must have votes of trust from at least half of the anchors in the common neighborhood.

Note that One could view DRBTS as an algorithm that allows a sensor to ignore location information from malicious anchors, enabling robust estimation of its location. But it could be also viewed as an algorithm that validates location claims of a particular anchor.

### 4.2.7 Summary

We surveyed schemes used for robust location estimation. In [11] a sensor node receives location information from a set of locators. Some resolution algorithms, such as ACLA, are used to detect attacks.

As an improvement, [19] achieves better accuracy for the location at the cost of more communication and computation. It is assumed that locators are able to vary their transmission ranges and transmission directions.

In [20] a node collects distances to a set of locators. After that, one of two methods may be used to filter out malicious locators. One method is attack-resistant MMSE which filters out locators according to the squared error they contribute to the mean square error. Another method is voting based location estimation, in which the search space is divided into a grid. Then, a voting is made on each grid cell to determine its goodness as a location estimate. The centroid of the cells with the highest vote is used as an estimate.

In [13] special integrity-codes are used to encode messages broadcasted by locators. These integrity codes allow nodes to easily detect many attacks.

In [14] locators are assumed to have a directional antenna. If a sensor lies in a triangle formed by three locators and each can do distance bounding with the sensor, verifiable multilateration is used to estimate its location. Otherwise, given the location of each nearby locator, the sensor calculates the region of interest as the intersection of two regions. First, is discs formed by transmission beam of nearby locators. Second, region determined by distance bounds obtained in the earlier phase.

DRBTS [22] uses the concept of reputation to eliminate the malicious anchors. Each anchor node maintains a Neighbor-Reputation-Table (NRT) and a sensor node will use the location information of an anchor $B$ only if the majority of anchors in the vicinity trust $B$.

# 5 Discussion

| Algorithm | Classification | | | | |
|---|---|---|---|---|---|
| | Objective | Encryption keys | Signal frequency | Use of Anchors | Range based/free |
| Distance bounding protocol [6] | Verify claims | Used | RF | Free | Based |
| Secure verification of location claims [4] | Verify claims | Optional | Both RF & ultrasound | Free | Based |
| Secure positioning of wireless devices [16] | Verify claims | Used | RF | Free | Based |
| SLA [17] | Verify claims | Used | RF | Based | Based |
| Location verification using secure distance-bounding protocol [7] | Verify claims | Used | RF | Free | Based |
| A low-cost robust localization scheme for WLAN [18] | Verify claims | Used | RF | Based | Based |
| SecNav [13] | Own localization | Not used | RF | Based | Free |
| SeRLoc [11] | Own localization | Used | RF | Based | Based |
| HiRLoc [19] | Own localization | Used | RF | Based | Based |
| Attack-resistant location estimation in sensor networks [20] | Own localization | Not used | RF | Free | Free |
| ROPE [14] | Own localization | Used | RF | Based | Based |
| DRBTS [22] | Own localization | Used | RF | Based | Based |

Table 1: Applying the classification to secure localization algorithms

Table 1 gives a comparison between the secure localization algorithms discussed in the paper. Although the two objectives "verify location claims" and "robust location estimation" are not contradicting, none of the discussed algorithms gives a framework that handles both objectives. All protocols in this survey make use of RF signals. RF signals disallow an adversary to shorten measured distances because it propagates at the speed of light. However, it needs very accurate calculations and measurements because small errors in calculations produce large distance errors. Only "Secure verification of location claims" also makes use of ultra-sound, which makes it vulnerable to physically present attacks. Ultrasound is better in that it can give accurate results when there is an error in measurements. All protocols under "robust location estimation" are anchor-based, except for "Attack-resistant location estimation in sensor networks" in which each node collects distances to other nodes. When there are anchors in the network, they allow for more accurate results because their locations are found using GPS. Also some algorithms assume that anchors are protected from forgery. This allows for easier algorithms and simpler systems designs.

| Algorithm | Localization Attacks ||||||||||
|---|---|---|---|---|---|---|---|---|---|---|
| | Distance Fraud | Mafia Fraud | Terrorist Fraud | Wormhole | Sybil | Spoofing | Jamming | Overshadowing | Manipulation | Replay |
| Distance bounding protocol [6] | ■ | ■ | | ■ | | | | | | |
| Secure verification of location claims [4] | ■ | | | ■ | | | | | | |
| Secure positioning of wireless devices [16] | ■ | ■ | | | | | | | | |
| SecNav [13] | | | | | | ■ | ■ | ■ | ■ | ■ |
| SLA [17] | ■ | | | | | | | | | ■ |
| SeRLoc [11] | | | | ■ | ■ | | | | | |
| HiRLoc [19] | | | | ■ | ■ | | | | | |
| Location verification using secure distance-bounding protocol [7] | ■ | ■ | ■ | | | | | | | |
| A low-cost robust localization scheme for WLAN [18] | ■ | | | | | ■ | | | ■ | |
| Attack-resistant location estimation in sensor networks [20] | | | | | ■ | ■ | | | | |
| ROPE [14] | ■ | | | | | ■ | | | | |
| DRBTS [14] | ■ | | | ■ | | ■ | | | | |

Table 2: Summary of security attacks addressed by each algorithm

Verification of location claims is done at three granularities: "distance-bounding" verifies a distance claim, "secure verification of location claims" verifies an in-region claim (i.e. verifies that a prover lies within a specified region), while "verifiable multilateration" verifies that a prover is at a particular location.

All discussed protocols under "verify location claims" can make use of encryption keys, which makes authentication easier and more secure. They are all range-based. Yet, "SLA" and "A low-cost robust localization scheme for WLAN" are classified as range-free when they use message-maps. Range-free algorithms are less accurate when compared to range-based ones. On the other hand, they are scalable because they do not count on special types of nodes.

Table 2 shows which localization algorithms address which security attacks. By definition, all discussed protocols under "verify location claims" target the distance fraud attack. Among these, [6], [16] and [7] also target the mafia fraud attack, while only [7] targets the terrorist fraud attack. Wormhole and spoofing (malicious node pretending to be privileged) attacks got the attention of many algorithms. Both [13] and [18] detect manipulation of localization messages by malicious nodes. SecNav [13] is the only localization algorithm in WSN -to the best of our knowledge- that targets jamming, overshadowing, and replay attacks.

# 6 Conclusion and Open Issues

In this survey paper, we have illustrated the importance of securing the localization process in wireless sensor networks. We presented a taxonomy of the different features that can be used to classify secure localization schemes for wireless sensor networks. The taxonomy can be applied to evaluate the characteristics of a location system needed by a particular application or the suitability of an existing location system for its application. Further, attacks against localization schemes in WSNs were explained. Then, secure localization schemes suitable (or potentially suitable) for sensor networks are discussed, mentioning the attacks each of them targets.

Many research opportunities are still available in the field. Protocols that make use of ultra-sound in clever ways to secure the localization process are missing in the literature. Algorithms tailored for WSN which verify location claims without requiring a point-to-point range estimates (i.e. range-free) are missing. Range-free algorithms are more scalable because it does not need special types of nodes like anchors which is more suitable for WSNs. For WSN applications where both objectives (i.e. verify claims & own localization) are needed, a unified framework that combines mechanisms for solving both problems will be appreciated. A whole new objective in the field is to hide a legitimate sensor's location (e.g. for privacy concerns) from other nodes (e.g. an opponent), while enabling the sensor to transmit and receive messages. Protocols that enable a group of nodes to know their locations without giving opponent nodes enough clues would be of interest for many civil and military applications. Secure localization protocols that account for movement of nodes are lacking. Imagine an application in which nodes are allowed to move along specific paths. It is required to periodically verify that nodes are still on track.

Although secure localization in WSNs is getting researchers' attention recently, it is still a "young field"; meaning, the door is all-the-way open for new ideas to enrich the literature. We hope this survey paper could encourage more research in the field and drive attention to new/open issues in the field.